\documentclass[a4paper,12pt]{article}
\pdfoutput=1
\usepackage{epsfig}
\usepackage{amssymb}
\usepackage{amsfonts}
\usepackage{amsmath}
\usepackage{amsthm}
\usepackage{euscript}
\usepackage{verbatim}
\usepackage{latexsym}
\usepackage{graphicx}
\usepackage{caption}
\usepackage{float}
\usepackage{braket}
\usepackage{subcaption}
\usepackage{enumitem}
\usepackage{comment}
\usepackage{tensor}
\usepackage{listings}
\usepackage{slashed}

\usepackage{mathrsfs}

\newif\ifdtup

\catcode`\@=11

\@addtoreset{equation}{section}

\def\@normalsize{\@setsize\normalsize{15pt}\xiipt\@xiipt
\abovedisplayskip 14pt plus3pt minus3pt%
\belowdisplayskip \abovedisplayskip
\abovedisplayshortskip \z@ plus3pt%
\belowdisplayshortskip 7pt plus3.5pt minus0pt}

\def\small{\@setsize\small{13.6pt}\xipt\@xipt
\abovedisplayskip 13pt plus3pt minus3pt%
\belowdisplayskip \abovedisplayskip
\abovedisplayshortskip \z@ plus3pt%
\belowdisplayshortskip 7pt plus3.5pt minus0pt
\def\@listi{\parsep 4.5pt plus 2pt minus 1pt
     \itemsep \parsep
     \topsep 9pt plus 3pt minus 3pt}}

\relax

\catcode`@=12

\catcode`\@=11

\def\section{\@startsection{section}{1}{\z@}{3.5ex plus 1ex minus
   .2ex}{2.3ex plus .2ex}{\large\bf}}

\def\SymBoxes#1#2#3#4{\newdimen\un@t \un@t#3%
\raisebox{#1}{\rule{#2\un@t}{#4}\hskip-#2\un@t
\@tempdimb\un@t \advance\@tempdimb by-#4\@tempcntb#2\relax%
\@whilenum{\@tempcntb>0}\do{
\rule{#4}{\un@t}\hskip\@tempdimb \advance\@tempcntb by\m@ne}%
\hskip-#2\un@t \rule[\un@t]{#2\un@t}{#4}%
\rule[\un@t]{#4}{#4}\hskip-#4
\rule{#4}{\un@t}}\hskip-#4}                

\begin{document}

\newcommand{\beq}{\begin{equation}}
\newcommand{\eeq}{\end{equation}}
\newcommand{\bea}{\begin{eqnarray}}
\newcommand{\eea}{\end{eqnarray}}
\newcommand{\beas}{\begin{eqnarray*}}
\newcommand{\eeas}{\end{eqnarray*}}
\newcommand{\defi}{\stackrel{\rm def}{=}}
\newcommand{\non}{\nonumber}
\newcommand{\bquo}{\begin{quote}}
\newcommand{\enqu}{\end{quote}}
\renewcommand{\(}{\begin{equation}}
\renewcommand{\)}{\end{equation}}
\def \eqn#1#2{\begin{equation}#2\label{#1}\end{equation}}

\def\e{\epsilon}
\def\IZ{{\mathbb Z}}
\def\IR{{\mathbb R}}
\def\IC{{\mathbb C}}
\def\IQ{{\mathbb Q}}
\def\de{\partial}
\def\Tr{ \hbox{\rm Tr}}
\def\H{ \hbox{\rm H}}
\def\HE{ \hbox{$\rm H^{even}$}}
\def\HO{ \hbox{$\rm H^{odd}$}}
\def\K{ \hbox{\rm K}}
\def\Im{ \hbox{\rm Im}}
\def\Ker{ \hbox{\rm Ker}}
\def\const{\hbox {\rm const.}}
\def\o{\over}
\def\im{\hbox{\rm Im}}
\def\re{\hbox{\rm Re}}
\def\bra{\langle}\def\ket{\rangle}
\def\Arg{\hbox {\rm Arg}}
\def\Re{\hbox {\rm Re}}
\def\Im{\hbox {\rm Im}}
\def\exo{\hbox {\rm exp}}
\def\diag{\hbox{\rm diag}}
\def\longvert{{\rule[-2mm]{0.1mm}{7mm}}\,}
\def\a{\alpha}
\def\dag{{}^{\dagger}}
\def\tq{{\widetilde q}}
\def\p{{}^{\prime}}
\def\W{W}
\def\N{{\cal N}}
\def\hsp{,\hspace{.7cm}}

\def\br{\nonumber}
\def\IZ{{\mathbb Z}}
\def\IR{{\mathbb R}}
\def\IC{{\mathbb C}}
\def\IQ{{\mathbb Q}}
\def\IP{{\mathbb P}}
\def \eqn#1#2{\begin{equation}#2\label{#1}\end{equation}}

\newcommand{\C}{\ensuremath{\mathbb C}}
\newcommand{\Z}{\ensuremath{\mathbb Z}}
\newcommand{\R}{\ensuremath{\mathbb R}}
\newcommand{\rp}{\ensuremath{\mathbb {RP}}}
\newcommand{\cp}{\ensuremath{\mathbb {CP}}}
\newcommand{\vac}{\ensuremath{|0\rangle}}
\newcommand{\vact}{\ensuremath{|00\rangle}                    }
\newcommand{\oc}{\ensuremath{\overline{c}}}
\newcommand{\psizero}{\psi_{0}}
\newcommand{\phizero}{\phi_{0}}
\newcommand{\hzero}{h_{0}}
\newcommand{\psiin}{\psi_{\rh}}
\newcommand{\phiin}{\phi_{\rh}}
\newcommand{\hin}{h_{\rh}}
\newcommand{\rh}{r_{h}}
\newcommand{\rb}{r_{b}}
\newcommand{\psibnd}{\psi_{0}^{b}}
\newcommand{\psibndp}{\psi_{1}^{b}}
\newcommand{\phibnd}{\phi_{0}^{b}}
\newcommand{\phibndp}{\phi_{1}^{b}}
\newcommand{\gbnd}{g_{0}^{b}}
\newcommand{\hbnd}{h_{0}^{b}}
\newcommand{\zh}{z_{h}}
\newcommand{\zb}{z_{b}}
\newcommand{\man}{\mathcal{M}}
\newcommand{\hbr}{\bar{h}}
\newcommand{\tbr}{\bar{t}}
\newcommand{\zbr}{\bar{z}}
\newcommand{\wbr}{\bar{w}}

\newcommand{\tlam}{\tilde\lambda}
\newcommand{\tOm}{\tilde\Omega}
\newcommand{\tC}{\tilde{\mathcal{C}}}
\newcommand{\tD}{\tilde{\mathcal{D}}}
\newcommand{\tE}{\tilde{\mathcal{E}}}
\newcommand{\tal}{\tilde\alpha}

\newcommand{\scrip}{\mathscr{I}^{+}}
\newcommand{\scrim}{\mathscr{I}^{-}}
\newcommand{\scri}{\mathscr{I}}

\begin{titlepage}
\begin{flushright}
CHEP XXXXX
\end{flushright}
\bigskip
\def\thefootnote{\fnsymbol{footnote}}

\begin{center}
{\Large
{\bf Soft Hair on Schwarzschild: \\
\vspace{0.1in}
A Wrinkle in Birkhoff's Theorem
}
}
\end{center}

\bigskip
\begin{center}
Chethan KRISHNAN$^a$\footnote{\texttt{chethan.krishnan@gmail.com}}, \ Jude PEREIRA$^b$ \footnote{\texttt{jude.pereira@asu.edu}} 
\vspace{0.1in}

\end{center}

\renewcommand{\thefootnote}{\arabic{footnote}}

\begin{center}
\vspace{-0.2cm}

$^a$ {\it Center for High Energy Physics, Indian Institute of Science, \\ C V Raman Road, Bangalore 560012, India}\\
\vspace{0.2cm}
$^b${\it Department of Physics, Arizona State University,\\
Tempe, Arizona 85287-1504, USA.}
\end{center}



\noindent
\begin{center} {\bf Abstract} \end{center}
The double null form of the Schwarzschild metric is usually arrived at by demanding Eddington-Finkelstein (EF) conditions at the horizon. This leads to certain logarithmic fall-offs that are too slow along null directions at $\scri$, resulting in divergences in the covariant surface charges. These coordinates are therefore {\em not} asymptotically flat. In this paper, we find a natural alternative double null form for Schwarzschild that is adapted to $\scrip$ or $\scrim$  instead of the horizon. In its final form, the metric has only power law fall-offs and fits into the recently introduced Special Double Null (SDN) gauge, with finite surface charges. One remarkable feature of SDN gauge is that  spherical symmetry and vacuum Einstein equations allow an infinite number of asymptotic integration constants in the metric, on top of the mass. This is an apparent violation of Birkhoff's theorem. We note however that all except two of these new parameters are absent in the charges, and therefore correspond to trivial hair. The remaining two parameters do show up in the charges, depending on the choice of allowed fall-offs. We provide an understanding of this observation -- Birkhoff's theorem fixes Schwarzschild only {\em up to diffeomorphisms}, but diffeomorphisms need not vanish at infinity and can in principle become global symmetries. If such asymptotic diffeomorphisms are spherically symmetric, their associated soft modes can become Birkhoff hair.  The relevant global symmetries here are certain hypertranslation shifts in the $v$-coordinate at $\scrip$ (and $u$ at $\scrim$), which are inaccessible in other gauges.

\vspace{1.6 cm}
\vfill

\end{titlepage}

\setcounter{footnote}{0}


\section{Introduction}

Birkhoff's theorem\footnote{The theorem \cite{wikiBirkh} is more correctly referred to as the Jebsen-Birkhoff theorem -- Jorg Jebsen (1888-1922) had it in 1921, two years before Birkhoff \cite{Jebsen, Birkhoff}.} states that if we assume spherical symmetry, there is a unique one-parameter family of solutions to the vacuum Einstein equations
\bea
R_{\mu\nu}=0. \label{vacE}
\eea
This is the famous Schwarzschild metric, which contains a single parameter $M$. Spherical symmetry reduces the number of unknown functions in the metric to two \cite{Carroll}. It also guarantees that these functions do not depend on the sphere coordinates. Therefore the way Birkhoff's theorem is usually arrived at, is in two steps. First, one writes down a metric with two unknown functions of the two (non-sphere) coordinates. Second, we solve the components of \eqref{vacE} in a clever sequence \cite{Stackexchange}. The result is the Schwarzschild metric, with the mass parameter $m$ showing up as an integration constant in one of the steps. The surprising feature of Birkhoff's theorem is that we have assumed neither that the spacetime is asymptotically flat\footnote{The term ``asymptotic flatness'' here is to be understood loosely. It simply means that the leading behavior of the metric at large $r$ is that of Minkowski, where $r$ and $t$ are the usual Schwarzschild coordinates. Our paper can be viewed as an exploration of the precise notion of asymptotic flatness relevant for understanding Birkhoff's theorem.}, nor that it is static. The above strategy forces both of these features upon us.

Of course, to implement the above two-step strategy explicitly, we have to make a choice of coordinates. The choice that one usually works with is the standard $t$ and $r$ (and angle) coordinates. In this paper, we will start with a small technical observation -- the success of the second step above relies on the choice of these coordinates. In particular, it fails when we work with a double null ($u$ and $v$) coordinate system instead of $t$ and $r$. The precise nature of this failure and its interpretation will be a major topic of this paper.  One of our main messages is that changes of coordinates (diffeomorphisms) can become physical in the asymptotic region of spacetime, and in double null coordinates, they can lead to  spherically symmetric soft hair parameters in the metric. A closely related idea is in fact familiar from Bondi gauge. One can do a supertranslation on the Schwarzschild metric to obtain a solution that is diffeomorphic to Schwarzschild, but with supertranslation hair (see eg. \cite{StromReview, Compere, Takeuchi}). But there is a difference  -- if we insist on spherical symmetry, the only spherically symmetric supertranslation is an ordinary time translation which leaves the Schwarzschild metric invariant. Therefore in Bondi gauge, Schwarzschild has no spherically symmetric hair even if we allow supertranslations. The crucial new ingredient in the present paper is that we work with a double null coordinate system (to be elaborated below) and this leads to new diffeomorphisms and soft hair on Schwarzschild that can be {\em spherically symmetric}. This is a new wrinkle in Birkhoff's theorem, and explains the subtitle of this paper. 

Our motivations for investigating this problem have their origins beyond Birkhoff's theorem. Our initial goal was to try and understand the Schwarzschild metric in the recently introduced Special Double Null (SDN) gauge \cite{KP, KP1, KP2} (see also related work in \cite{ACD, Page}). What we initially believed would be a small exercise, turned out to be a somewhat tricky question tying together various subtleties involving diffeomorphisms, global symmetries, bulk isometries, fall-offs and surface charges. 
These questions eventually lead us to a new understanding of Birkhoff.

Asymptotically flat spacetimes are usually studied in settings where the ``holographic'' direction is a radial spacelike direction \cite{Bondi, Ashtekar}. SDN gauge was introduced as a potentially useful setting for studying flat space, in which $\scrip$ and $\scrim$ are approached along {\em null} directions instead. Various motivations for this choice are given in \cite{KP,KP1,KP2}, see especially the introductions of \cite{KP,KP2}. In the first paper \cite{KP}, we introduced the SDN gauge 
and studied some of its new features -- the observation that power law fall-offs along null directions are sufficient to allow black holes, and the existence of whole towers of trivial diffeomorphisms, are two such features that will play key roles in the present paper. 
In \cite{KP1, KP2}, it was further shown that a suitably defined asymptotic symmetry algebra in the SDN gauge can be significantly bigger than the BMS algebra \cite{Bondi}. This bigger algebra will not directly play a role in our discussions here, but the hair associated to these extra diffeomorphisms (hypertranslations and their cousins \cite{KP1,KP2}) will be important in our story. 

Writing Schwarzschild metric in the SDN gauge as an asymptotic solution can be done straightforwardly enough, by solving the Einstein equations order by order in the power law fall-offs considered in \cite{KP, KP1} while restricting to spherical symmetry. The non-trivial problem is to construct a closed form -- by which we mean a definition of the Schwarzschild metric that is non-perturbative in the $\frac{1}{v}$ fall-offs at $\scrip$ (or in $\frac{1}{u}$ at $\scrim$ for the past chart). Considering the fundamental significance of the Schwarzschild solution, finding a closed form definition\footnote{Let us emphasize that by ``closed form'' we do {\em not} mean the ability to express the solution in terms of simple functions. This is not possible even for Schwarzschild in the conventional double null gauge or Kruskal coordinates. The coordinates there are defined implicitly. But these do constitute complete non-perturbative definitions. We will see that an exactly analogous situation arises for Schwarzschild in the SDN gauge as well -- our non-perturbative definition will also be in terms of implicit functions. Explicit forms of the coordinate transformation can be found in Section \ref{Final}, see \eqref{major} for the general form and \eqref{simplest} for a simple choice.} for it, can be expected to  be instructive. We will find that this is indeed the case, with an interesting interplay between various features.  
We will also find that the conventional form of the Schwarzschild metric in double null coordinates (see eg. \cite{SS} or our eqn \eqref{dnS}), obtained  by defining a double Eddington-Finkelstein coordinate system at the {\em horizon}, is unsatisfactory for a number of reasons. Most crucially, we will show that this coordinate system is technically not asymptotically flat, and therefore not ideal for the study of holography and asymptotic symmetries. We will find a better way to define Schwarzschild in double null coordinates, and this will naturally be consistent with the fall-offs considered in \cite{KP2}. We will clarify the relationship to Birkhoff's theorem alluded to in the beginning of this section and the eventual final form of the metric presented in Section \ref{Final}.


We will organize the paper by first presenting the most pedestrian attempt to construct Schwarzschild in a double null coordinate system. As the subtleties reveal themselves, we will become more systematic and formal. We feel that this is the most transparent way to arrange our results; starting with the generalities of the SDN gauge (which is how we arrived at this question) makes the results seem more surprising than they are. A summary of curiosities is presented in the Conclusions section, after the dust settles. A reader not interested in the details can proceed directly there.

\section{First Pass: The Schmidt-Stewart Metric}

We start by pointing out that there is a perfectly plausible candidate for a double null metric for Schwarzschild. This is the ``double'' Eddington-Finkelstein coordinates that one can arrive at by starting with conventional Schwarzschild, and switching to Eddington-Finkelstein coordinates in {\em both} the past and the future:
\bea
u=t-r-2m \ln |r-2m|, \ \ 
v=t+r+2m \ln |r-2m|
\eea
This leads to a metric
\bea
ds^2= -\Big(1-\frac{2m}{r(u,v)}\Big)\, du\ dv + r^2(u,v) \ d\Omega_2^2 \label{dnS}
\eea
where $r(u,v)$ is to be understood as being implicitly solved via\footnote{Note again that this provides a non-perturbative definition of the metric, but the definition is implicit.} 
\bea
\frac{v-u}{2}=r+2m \ln |r-2m| \label{rvminusu}
\eea
This form of the metric was probably known for some time\footnote{But note that the origin of even the conventional Eddington-Finkelstein metric is uncertain \cite{WikiEddFink}.}, but Schmidt and Stewart studied it in detail in an old, interesting and largely ignored paper \cite{SS}. They made various thought-provoking observations about solutions of scalar wave equations on the Schwarzschild black hole in this paper, which we feel are still of interest from a more modern (holographic) perspective. To make belated amends, we will call \eqref{dnS} with \eqref{rvminusu}, the Schmidt-Stewart metric.

Note that the metric \eqref{dnS} satisfies the SDN gauge conditions \cite{KP} 
\bea
g^{uu}=g^{vv}=0,  \ \ g^{uA}=g^{vA}.
\eea 
But we also need to ensure that it satisfies suitable fall-off requirements. In what follows, we will typically describe the future chart around $\scrip$ where $v\rightarrow \infty$, but similar statements apply for $\scrim$ as well where $u \rightarrow -\infty$. In \cite{KP, KP1, KP2}, we considered various power-law fall-offs. We also argued that power law fall-offs were sufficient to incorporate Kerr black holes and gravitational radiation. 

Can one write   \eqref{dnS} at large-$v$ as a power series in $\frac{1}{v}$? If that can be done, then the Schmidt-Stewart metric can be a viable candidate for the SDN gauge form of Schwarzschild.  
To check if this is true, let us first invert the above expression \eqref{rvminusu} to write 
\bea\label{1byrSS}\begin{aligned}
\frac{1}{r}=&\frac{2}{v-u}+\frac{8m}{(v-u)^2} \ln (v-u)+\frac{32m^2}{(v-u)^3} [\ln (v-u)]^2-\frac{32m^2}{(v-u)^3} \big[\ln (v-u)\big]\\&-\frac{32m^2}{(v-u)^3}+\mathcal{O}\bigg(\frac{\ln (v-u)]^3}{(v-u)^4}\bigg)
\end{aligned}\eea
The fist two terms in this expansion have been noted earlier in \cite{SS}. We can further expand the right hand side by choosing the conformal boundary that we are interested in, and we pick $\scrip$ for concreteness. This means that $v \rightarrow \infty$. Writing the above expression as a double expansion in powers of $1/v$ and $\ln v$, 
\bea
\ln \left(v-u\right)= 
 \ln v - \left(\frac{u}{v}+\frac{u^2}{2v^2}+\cdots\right)
\eea
we get the result
\bea\label{FullSchmidt}
\begin{aligned}
\frac{1}{r} &= \frac{2}{v}+\frac{8m}{v^2}\, \ln v+\frac{2(u-4m\ln 2)}{v^2}+\frac{32m^2}{v^3}(\ln v)^2+\frac{16m(u-4m\ln 2-2m)}{v^3}\ln v \\&+\frac{\big(2u^2-8mu-32m^2+32m^2(\ln 2)^2-16mu\ln 2+32m^2\ln 2\big)}{v^3}+\mathcal{O}\bigg(\frac{(\ln v)^3}{v^4}\bigg)
\end{aligned}
\eea
A key consequence of this expression and the eventual Schmidt-Stewart metric resulting from it, is that it contains fall-off terms which are logarithmic in $v$. In other words, it does not fall into the power law fall-offs discussed in \cite{KP, KP1, KP2}. Though a discussion of log fall-offs was not presented in \cite{KP, KP1, KP2} because physically interesting spacetimes can already be included with power laws, it is indeed possible to consider a double expansion in both $\log v$ and $1/v$. A general treatment of this was done in \cite{KPBig} which is yet to appear on the arXiv. Here we will only discuss  some less technical aspects of a baby-version for the spherically symmetric case, see Appendices \ref{logs} and \ref{logdiffs}.  It can be checked using the formulas for the charges in \cite{Wald, Brandt}  (see also the final Appendices in \cite{KP1, KP2}) that most log terms lead to charge contributions that vanish and therefore correspond to trivial hair. If this were universally true, one could view Schmidt-Stewart as simply a trivial coordinate transformation away from the SDN gauge power law fall-offs considered in \cite{KP, KP1, KP2}. 

But there are two flies in the ointment. Firstly, log terms are generally an unpleasant feature. Having log terms makes the expansion non-analytic at infinity (here $\scrip$). See eg. \cite{Kehrberger} for some recent discussions and references on related matters in Bondi gauge.  Secondly, the large-$v$ behavior of Schmidt-Stewart metric that follows from \eqref{FullSchmidt} is not merely unpleasant, some of the terms in it lead to divergences in the charges. We will not present the details here, but if one computes the charges using the formulas in  \cite{Wald, Brandt, KP1, KP2}, one finds that if the term $\lambda_{01}$ (see eqn. \eqref{loglambdaSch}) in the metric is turned on, there are divergences proportional to $\lambda_{01}$ that go as ${\cal O}(v \log v)$. It can be checked that the Schmidt-Stewart metric in the SDN gauge arising from \eqref{FullSchmidt} indeed contains the $\lambda_{01}$ term and therefore is problematic. This can also be seen by noting that the formulas in Appendices \ref{logs} and \ref{logdiffs} show that a non-trivial $r_{12}$ term  in the general asymptotic coordinate transformation
\eqref{rtransformSch} can generate the $\lambda_{01}$ term -- it is easy to see that \eqref{FullSchmidt} fits into the form \eqref{rtransformSch} with a non-vanishing $r_{12}$ term.


Divergent charges imply that the Schmidt-Stewart metric is technically {\em not} asymptotically flat. But let us emphasize that not all log terms in the metric expansion lead to divergent surface charges. In fact, it is possible to generalize the fall-offs considered in \cite{KP, KP1, KP2}, so that one includes certain infinite classes of log terms \cite{KPBig} which correspond to trivial diffeomorphisms. But the log terms present in the Schmidt-Stewart metric arise too early in the expansion and lead to divergences. In particular, it generates $\lambda_{01}$ (the analogue of $\lambda_1$ in \eqref{polylambdaSch}, but when there are logs as well) which is forced to be zero by the Einstein equations if only power law fall-offs are allowed. As mentioned already, 
from our work in the previous papers (see also \cite{KPBig}) we know that this is sufficient to capture all the solutions in the Bondi class, including Schwarzschild, Kerr and radiative data (like shear and news). 




\section{Schwarzschild in SDN Gauge: Asymptotic Definition}\label{SDNexp}

Given that the Schmidt-Stewart form is not suitable for discussing the asymptotic region, how are we supposed to construct Schwarzschild in SDN gauge? One way to proceed, is to use spherical symmetry to restrict the form of the metric. We start with with the most general spherically symmetric metric in the SDN gauge of \cite{KP, KP1}:
\beq 
ds^2=-e^{\lambda(u,v)}\, du\, dv +2\, \Big(\frac{v-u}{2}\Big)^2\, \Omega(u,v)\, \gamma_{z\zbr}\, dz\, d\zbr 
\eeq
We can plug this into the vacuum Einstein equations \eqref{vacE} and we will be left with a set of partial differential equations in $u$ and $v$. These are simple to write down, and are presented in Appendix \ref{SphericalEinstein}. Unlike in the $r$-$t$ coordinate system, where the system is easy to solve, here the situation is more complicated. We will see eventually that there are conceptual reasons for this, and will be able to find useful solutions. But for the moment, we will proceed  pragmatically and make further demands on the metric to make the system tractable. 

The first thing we could try, is to ask whether we can find a solution by demanding that the metric is a solution only of $(v-u)$ and not  $v$ and $u$ separately. This can be viewed as the explicit demand that the spacetime is stationary\footnote{In this paper by ``stationary", we will mean time-translation invariant. A spacetime is ``static", if it is stationary and also time-reversal/CPT invariant (ie., invariant under $u \leftrightarrow -v$). A spherically symmetric double null metric that is only dependent on $(v-u)$ is manifestly static.}. This is usually viewed as a consequence of Birkhoff's theorem and was true for the Schmidt-Stewart metric, so clearly solutions of this type exist. Once we make the assumption that the metric is only a function of the $(v-u)$ the equations satisfied by $\Omega$ and $\lambda$ become ODEs, these are also presented in Appendix \ref{SphericalEinstein}. A key point about these equations however is that if we demand that the solutions have a series expansion at large $(v-u)$ in powers of $1/(v-u)$ (ie., without $ \log$ terms), then the {\em only} solution is Minkowski! This is easy to check from the ODEs we write down in the Appendix. Note that this is consistent with the existence of Schmidt-Stewart metric, which contains log fall-offs, see \eqref{1byrSS} and \eqref{FullSchmidt}. Since we would like to avoid logs for reasons mentioned earlier, we will drop the assumption that the metric is only a function of $(v-u)$.\footnote{This also means that the question of whether the spacetime is stationary and static will be more subtle -- we will find that the spacetime is indeed static, but this is realized in a fairly remarkable way.}

Instead, we will ask that the metric is asymptotically flat with power law fall-offs in $\frac{1}{v}$ at $\scrip$. As pointed out in the Introduction, asymptotic flatness is not a premise of Birkhoff's theorem, it is usually viewed as a consequence. So by adding a specific form of asymptotic flatness as part of the premise (and as long as the fall-off demand is not too stringent), we expect to recover Schwarzschild. Moreover, since Birkhoff's theorem says that the only ``hair" allowed under the assumption of spherical symmetry is the mass parameter, it is reasonable to expect that we should at most get one independent integration constant in SDN gauge as well. But surprisingly, we will see that the form of the metric that results upon solving Einstein equations asymptotically, contains an infinite number of parameters on top of the mass. Somehow it seems that by explicitly adding asymptotic flatness as a premise, we have obtained a {\em more} general result! Interpreting these parameters will take us the rest of the paper, but first let us write down the perturbative (in $1/v$) solution.

We will take the fall-off conditions for the gauge functions $\lambda(u,v)$ and $\Omega(u,v)$ as  
\begin{subequations}
\begin{align}
    \label{polylambdaSch}
    \lambda(u,v) &= \frac{\lambda_{1}(u)}{v}+\frac{\lambda_{2}(u)}{v^2}+\frac{\lambda_{3}(u)}{v^3}+O\big(v^{-4}\big)\\
    \label{polyomegaSch}
    \Omega(u,v) &= 1+\frac{\mathcal{C}(u)}{v}+\frac{\mathcal{D}(u)}{v^2}+\frac{\mathcal{E}(u)}{v^3}+O\big(v^{-4}\big)
\end{align}    
\end{subequations}
An analogous expansion can be made for the past chart around $\scrim$ as well. 
Demanding that \eqref{polylambdaSch}-\eqref{polyomegaSch} satisfy Einstein's equations in vacuum allows us to solve the coefficients of the above expansion and the $u$-dependence of the coefficients get determined as polynomials. 
\begin{itemize}
    \item Constraints on $\Omega$ fall-offs in \eqref{polyomegaSch}:
    \begin{subequations}
        \begin{align}
        \partial_u\mathcal{C} &= 0 \implies \mathcal{C} = \text{constant} = \mathscr{C}\\
        \partial_u^2\mathcal{D} &= 0 \implies \mathcal{D} = \mathscr{D}^{(1)}\, u + \mathscr{D}^{(0)}\\
        \begin{split}
        \partial_u\mathcal{E} &= \big(3\, \mathscr{D}^{(1)}-\mathscr{C}\big)\, u+\Big(\mathscr{D}^{(0)}+\frac{1}{4}\, \mathscr{C}^2\Big) \\& \implies \mathcal{E} = \frac{1}{2}\, \big(3\, \mathscr{D}^{(1)}-\mathscr{C}\big)\, u^2+ \Big(\mathscr{D}^{(0)}+\frac{1}{4}\, \mathscr{C}^2\Big)\, u+\mathscr{E}
        \end{split}
        \end{align}
    \end{subequations}    
The coefficients on the right hand side are all constants.
    \item Constraints on $\lambda$ fall-offs in \eqref{polylambdaSch}:
    \begin{subequations}
        \begin{align}
        \lambda_{1} &= 0\\
        \label{polylambda2}
        \lambda_{2} &= \frac{1}{2}\, \big(\mathscr{C}-\mathscr{D}^{(1)}\big)\, u + \frac{1}{8}\, \big(\mathscr{C}^2-4\, \mathscr{D}^{(0)}\big)\\
        \lambda_{3} &= \frac{1}{2}\, \big(\mathscr{C}-\mathscr{D}^{(1)}\big)u^2-\frac{1}{2}\, \mathscr{C}\, \big(\mathscr{C}-\mathscr{D}^{(1)}\big) u-\frac{1}{8}\, \big(\mathscr{C}^3-4\, \mathscr{C}\, \mathscr{D}^{(0)}+8\, \mathscr{E}\big)
        \end{align}
    \end{subequations}
    Again the coefficients on the right hand side are constants.
\end{itemize}

\noindent
This constitutes the asymptotic definition of the Schwarzschild metric in SDN gauge and clearly we can extend this to arbitrarily high orders. We have presented some of the leading order terms. The key point about these asymptotic solutions is that at each order they generate new independent constants. This may seem superficially like an infinite violation of Birkhoff's theorem. 

One of our eventual punchlines will be that most of these integration constants have interpretations as trivial diffeomorphism hair. This means that they are just trivial coordinate transformations on the standard Schwarzschild metric. However, we will also see that the interpretation of three of these integration constants ($\mathscr{C}, \mathscr{D}^{(0)}, \mathscr{D}^{(1)}$) is more subtle. One combination of these three can be interpreted as the mass, but the other two do not have such a simple explanation. We could rule them out by fiat if we wish by choosing fall-offs that do not allow them, but we will need to resolve the puzzle of why they exist in the first place. More specifically, for the fall-offs chosen in \cite{KP2}, one can check that the charges (see \cite{KP1, KP2}) associated to these parameters are neither divergent nor vanishing. The general charge expression was written in \cite{KP2}, the explicit form of the covariant surface charge around the general spherically symmetric solution in SDN gauge is presented in Appendix \ref{charge}. Variations of both extra parameters show up in this charge. 

In the next section, we will first provide a brief description of what it means for a hair degree of freedom to be trivial or non-trivial. An understanding of this will also turn out to be essential for obtaining the non-perturbative definition of Schwarzschild that we are after. It is natural to suspect that there is some privileged choice of the trivial integration constants (perhaps setting them to zero) that will make the non-perturbative definition of the metric, simple. We will find that this is indeed the case.

Let us make one comment before we proceed, to connect our notation for the integration constants here with the notations we have used in \cite{KP, KP1, KP2}. Our demand of spherical symmetry has explicitly killed off the metric coefficients associated to supertranslations \cite{KP} and leading and subleading hyperrotations. These are the shear degrees of freedom $\mathcal{C}_{zz}, \mathcal{C}_{\zbr\zbr}$ and the integration ``constants" arising from $\alpha^A_2$ and $\alpha^A_3$ (see \cite{KP, KP1} for notations)\footnote{We put the ``constant'' in parenthesis, because in a general geometry they are arbitrary angle-dependent (but not $u$-dependent) integration ``constants". Of course, in the spherically symmetric situation that we are working with in this paper, they are true constants.}. Along with these, the integration ``constants'' discussed in \cite{KP, KP1, KP2} contain those relevant for hypertranslations and subleading hypertranslations ($\mathcal{C}_{z\zbr}$ and the second integration ``constant'' in $\lambda_2$  in the notations of \cite{KP, KP1, KP2}). These correspond to $\mathscr{C}$ and the $u$-independent term in \eqref{polylambda2}. Together with the first integration ``constant" of $\lambda_2$ which corresponds to the mass aspect in SDN gauge (the linear in $u$ term in \eqref{polylambda2}), we have therefore connected the three integration constants $\mathscr{C}, \mathscr{D}^{(0)}, \mathscr{D}^{(1)}$ to the relevant quantities in the more general discussion in \cite{KP1, KP2}. These three quantities will be responsible for the Schwarzschild mass and soft hair, as we will elaborate. 


\section{The Nature of Asymptotic Diffeomorphisms}

We will give a brief summary of the relationship between asymptotic symmetries and diffeomorphisms in gravitational theories in this section. The claims here are standard and reasonably widely known, but it is perhaps useful to summarize them because they will play a key role in our ensuing discussions. 

Diffeomorphisms are the gauge invariance of gravity and if one's goal is to ultimately build a quantum theory of gravity, then the observables of the theory have to be diff-invariant\footnote{In theories like string theory, in fact the gauge invariance of the description is expected to be bigger than diffeomorphisms.}. Crudely speaking, gauge transformations reduce to global symmetries at the asymptotic boundaries of spacetime. This is well-studied in the context of conventional gauge theories built from semi-simple Lie algebras. But the expectation is a natural one in gravity as well, where the true (holographic) observables are believed to be at the boundary. Indeed, this expectation is explicitly realized in the AdS/CFT correspondence \cite{Witten}. These observations suggest that asymptotic diffeomorphisms should have a natural interpretation in terms of the global symmetries of the quantum gravity Hamiltonian\footnote{This should be distinguished from the {\em bulk} global symmetries which are believed to be absent in quantum gravity. The global symmetries we have in mind should be compared to the conformal invariance or the $SO(6)$ R-symmetry of ${\mathcal N}=4$ SYM.}.   

But there are some nuances one should be careful about. To have a legitimate interpretation as a global charge with a well-defined action on the quantum gravity Hilbert space, the charges computed for these asymptotic diffeomorphisms should be both {\em non-vanishing} and {\em non-divergent}. The idea is that if the charges are vanishing, their actions do not change the state, whereas if they are divergent, they are not well-defined operators in the Hilbert space. So a minimal requirement in identifying the asymptotic symmetries of gravity is that the charges one computes around the states under question should be finite and non-vanishing. We will call such asymptotic charges {\em physical}, and view them as genuine global symmetries of the theory\footnote{We emphasize that identifying these asymptotic global symmetries of gravity is a somewhat tricky business, for many reasons. There are numerous choices on has to make including the boundary ($\scri$ or $spi$?), the gauge (Bondi, Ashtekar-Hansen or SDN?) and the allowed fall-offs (we want the allowed states to include black holes and radiative states, but even after this the choices are typically not unique). We will be working with the fall-offs considered in \cite{KP2} in this paper.}. A diffeomorphism that has a vanishing charge on the other hand is viewed  as {\em trivial}, it means that under its action the state does not change. 

In the papers \cite{KP1, KP2} we computed the {\em covariant} surfaces charges associated to the asymptotic diffeomorphisms present in SDN gauge (for two separate choices of allowed fall-offs). In more familiar Ashtekar-Hansen or Bondi gauges one can also work with {\em canonical} charges but the double null nature of SDN gauge makes it more natural to work with a covariant phase space formalism. The latter is developed in the work of Wald and collaborators \cite{Wald}, see also \cite{Brandt}. We have presented the general expressions for charges in \cite{KP1, KP2}. In an Appendix of this paper, we also present the form of the charges around the Schwarzschild background\footnote{In this paper, we will often use the word Schwarzschild to refer to the most general spherically symmetric solution in SDN gauge, which contains extra parameters above the mass.}. The parameters $\mathscr{C}, \mathscr{D}^{(0)}, \mathscr{D}^{(1)}$ show up in these charges, but the further subleading ones do not. The calculations that lead to these results also demonstrate that turning on the parameter $\lambda_1$ in \eqref{polylambdaSch} leads to divergent surface charges. This fact is related to the divergence of the charges in the Schmidt-Stewart coordinates, as we mentioned. This means that Schmidt-Stewart form of the double null metric cannot be interpreted as a state in the Hilbert space in which the ground state is the Minkowski vacuum in SDN gauge. 

The takeaway from the discussions of this section and the previous one is that all except  the three parameters $\mathscr{C}, \mathscr{D}^{(0)}, \mathscr{D}^{(1)}$ correspond to trivial hair, while these three parameters can be physical. We will see that $\mathscr{C}-\mathscr{D}^{(1)}$ is related to the mass parameter of the Schwarzschild black hole, while $\mathscr{C}$ and $\mathscr{D}^{(0)}$ are associated to broken hypertranslations and subleading hypertranslations. 

A further subtlety exists in the definition of physical charges, related to the appearance (or lack thereof) of the diffeomorphism (and not just the associated hair) in the charge expressions. This will be discussed in greater detail in a later section. We will loosely refer to $\mathscr{D}^{(0)}$ also as soft hair even though due to this subtlety, its interpretation has some ambiguities.

\section{Schwarzschild in SDN via Bondi}\label{viaBondi}

The first hint that the extra parameters we found in the SDN Schwarzschild metric are related to coordinate transformations near infinity, can be seen by connecting our results to Bondi gauge.  We begin with the Schwarzschild metric written in the Eddington-Finkelstein-Bondi coordinates $(u,r,z,\zbr)$:
\beq\label{SchwarzschildEF} ds^2= -\Big(1-\frac{2m}{r}\Big)\, du^2-2\, du\, dr+2\, r^2\, \gamma_{z\zbr}\, dz\, d\zbr\eeq
where $m$ is the mass of the Schwarzschild black hole. 
Next we transform the coordinate $r$ to write this metric in the SDN gauge. With the goal of connecting to both the allowed SDN gauge fall-offs as well as those arising in the Schmidt-Stewart metric, we consider a general expansion in both $1/v$ and $\log v$. 
\beq\label{rtransformSch}\begin{aligned}
\frac{1}{r}=& \bigg(\frac{r_{01}(u)}{v}+\frac{r_{02}(u)}{v^2}+\frac{r_{03}(u)}{v^3}+  \frac{r_{04}(u)}{v^4}+O\big(v^{-5}\big)\bigg)+(\ln{v})\bigg(\frac{r_{12}(u)}{v^2}+\frac{r_{13}(u)}{v^3}+\frac{r_{14}(u)}{v^4}+O\big(v^{-5}\big)\bigg)\\&+(\ln{v})^2\bigg(\frac{r_{23}(u)}{v^3}+\frac{r_{24}(u)}{v^4}+O\big(v^{-5}\big)\bigg)+(\ln{v})^3\bigg(\frac{r_{34}(u)}{v^4}+O\big(v^{-5}\big)\bigg)+O\big[(\ln{v})^4\big]
\end{aligned}\eeq
We have chosen the coefficients of our series expansion ansatz to depend exclusively on $u$. This ensures that the spherical symmetry of the metric is preserved under the coordinate transformation. Next we need to demand that the new metric obtained after substituting this ansatz for $r$ in \eqref{SchwarzschildEF} belongs to the SDN gauge. It is straightforward to check that the following conditions are automatically satisfied in the metric after the coordinate transformation \eqref{rtransformSch}, and therefore do not impose any restrictions on the $r_{ij}(u)$ coefficients:
\beq g^{uu}=0, \qquad g^{uA}=g^{vA}=0\eeq
However, we obtain constraints on the coefficients of \eqref{rtransformSch} from the SDN gauge condition $g^{vv}=0$. The solutions of these constraints for the first few orders are presented below:
\begin{subequations}
    \begin{align}
        \partial_u r_{01} &= 0 \implies r_{01} = \text{ constant } = \rho_{01}\\
        \partial_u r_{12} &= 0 \implies r_{12} = \text{ constant } = \rho_{12}\\
        \partial_u r_{02} &= \frac{1}{2}\, (\rho_{01})^2 \implies r_{02} = \frac{1}{2}\, (\rho_{01})^2\, u+\rho_{02}\\
        \partial_u r_{23} &= 0 \implies r_{23} = \text{ constant } = \rho_{23}\\
        \partial_u r_{13} &= \rho_{01}\, \rho_{12} \implies r_{13} = \rho_{01}\, \rho_{12}\, u + \rho_{13}\\
        \begin{split}
        \partial_u r_{03} &= \frac{1}{2}\, (\rho_{01})^3\, (u-2m)+\rho_{02}\, \rho_{01} \\& \implies r_{03} = \frac{1}{4}\, (\rho_{01})^3\, u^2 +\big(\rho_{02}\, \rho_{01}-m\, (\rho_{01})^3\big)\, u+\rho_{03}            
        \end{split}\\
        \partial_u r_{34} &= 0 \implies r_{34} = \text{ constant } = \rho_{34}\\
        \partial_u r_{24} &= \frac{1}{2}\, (\rho_{12})^2+\rho_{01}\, \rho_{23} \implies r_{24} = \frac{1}{2}\, (\rho_{12})^2\, u+\rho_{01}\, \rho_{23}\, u+\rho_{24}\\
        \begin{split}
        \partial_u r_{14} &= \frac{3}{2}\, \rho_{12}\, (\rho_{01})^2\, (u-2m)+\rho_{02}\, \rho_{12}+\rho_{13}\, \rho_{01} \\& \implies r_{14} = \frac{3}{4}\, (\rho_{01})^2\, \rho_{12}\, u^2 +\big(\rho_{13}\, \rho_{01}+\rho_{02}\, \rho_{12}-3\, m\, (\rho_{01})^2\, \rho_{12}\big)\, u+\rho_{14}            
        \end{split}\\
        \begin{split}
        \partial_u r_{04} &= \frac{3}{8}\, (\rho_{01})^4\, u^2+\frac{1}{8}\, \big(12\, \rho_{02}\, (\rho_{01})^2-20\, m\, (\rho_{01})^4\big)\, u+\frac{1}{8}\, \big(4\, (\rho_{02})^2+8\, \rho_{03}\, \rho_{01}-24\, m\, \rho_{02}\, (\rho_{01})^2\big) \\& \implies r_{04} = \frac{1}{8}\, (\rho_{01})^4\, u^3+\frac{1}{8}\, \big(6\, \rho_{02}\, (\rho_{01})^2-10\, m\, (\rho_{01})^4\big)\, u^2+\frac{1}{8}\, \big(4\, (\rho_{02})^2+8\, \rho_{03}\, \rho_{01} \\& -24\, m\, \rho_{02}\, (\rho_{01})^2\big)\, u+\rho_{04}
        \end{split}
    \end{align}
\end{subequations}

Thus we see that by simply demanding that the new coordinate transformed metric belongs to the SDN gauge, we are able to solve for the coefficients of the series expansion ansatz \eqref{rtransformSch} in terms of a set of constants $\{\rho_{01}, \rho_{12}, \rho_{02}, \rho_{23},\rho_{13},\rho_{03},\ldots\}$.  

Let us consider first the case where the log terms are set to zero. (We will discuss the log terms in more detail, in Appendices.) This means that all $\rho_{ij}$ with $i \ge 1$ are zero. Furthermore $\rho_{01}$ gets fixed to 2, because we want the leading behavior to be Minkowski (eg., we want the $\lambda$ expansion in \eqref{polylambdaSch} to $not$ start at $\mathcal{O}(v^0)$). It is straightforward to check that the remaining constants map exactly to the set of extra integration constants that we found in Section \ref{SDNexp} (which were obtained by solving the Einstein equations in vacuum assuming power law fall-offs).  By comparing the metric coefficients, we can in fact make this fully explicit. This shows that these extra parameters are all to be viewed as diffeomorphism hair associated to {\em spherically symmetric} diffeomorphisms. As we mentioned earlier, two of these extra parameters show up in the charges and therefore can have interpretation as genuine soft hair, while the rest are trivial parameters. We conclude this section by writing down the map between these non-trivial integration constants and the corresponding $\rho_{ij}$'s for the first few coefficients.
\begin{itemize}
    \item Constraints from comparing $g^{uv}$:
    \begin{subequations}
    \begin{align}
        \rho_{01} &= 2\\
        \rho_{12} &= 0\\
        \rho_{23} &= 0\\
        \rho_{13} &= 0\\
        \lambda_{02} &= -4\, m\, u - \frac{1}{4}\, (\rho_{02})^2+\frac{1}{2}\, \rho_{03}\\
        \rho_{34} &= 0\\
        \rho_{24} &= 0\\
        \rho_{14} &= 0\\
        \lambda_{03} &= -4\, m\, u^2-4\, \rho_{02}\, m\, u+\Big(\frac{1}{4}(\rho_{02})^3-\rho_{02}\, \rho_{03}+\rho_{04}\Big)
    \end{align}
    \end{subequations}
    \item Constraints from comparing $g^{z\zbr}$:
    \begin{subequations}
    \begin{align}
        \mathscr{C} &= -\rho_{02}\\
        \begin{split}
        \mathcal{D} &= \big(8\, m-\rho_{02}\big)\, u+\Big(\frac{3}{4}\, (\rho_{02})^2-\rho_{03}\Big)\\&
        \implies \mathscr{D}^{(1)} =8\, m-\rho_{02}\\&
        \implies \mathscr{D}^{(0)} =\frac{3}{4}\, (\rho_{02})^2-\rho_{03}
        \end{split}\\
        \mathscr{E} &= \frac{1}{2}\big(3\, \rho_{02}\, \rho_{03}-(\rho_{02})^3-2\, \rho_{04}\big)
    \end{align}
    \end{subequations}    
\end{itemize}
Note that the mass parameter is a linear combination of the integration constants:
\bea
m=\frac{\mathscr{D}^{(1)}-\mathscr{C}}{8}. 
\eea
The integration constants $\mathscr{C}, \mathscr{D}^{(0)}, \mathscr{D}^{(1)}$ therefore contain two extra pieces of data on top of the Schwarzschild mass.

\section{The Final Form}\label{Final}

Now that we have very explicitly identified the extra integration constants in SDN gauge as being associated to spherically symmetric diffeomorphisms, we  have a strategy for obtaining a closed form for Schwarzschild in the SDN gauge as follows. Instead of \eqref{rtransformSch}, which is an asymptotic coordinate transformation, we try 
\begin{eqnarray}
r \equiv r(u,v),
\end{eqnarray}
and then we demand the SDN gauge conditions. The non-trivial demand is $g^{vv}=0$ and remarkably, it turns out to be a simple differential equation which we can solve completely. The remarkable and simple result is 
\bea
2 r + 4 m \log |r-2m| = -u+f(v), \label{major}
\eea
where $f(v)$ is an arbitrary function of the $v$-coordinate. This is a very general definition of Schwarzschild in SDN gauge, but it is in some ways more general than what we would like. We are mostly interested in fall-offs at infinity that are power law\footnote{But we will also discuss the case with log fall-offs as well in Appendices and connect it to the Schmidt-Stewart metric as a special case.} in $1/v$. The function $f(v)$ is clearly a way of encoding all the parameters in the diffeomorphisms we encountered in the previous sections. If we are only seeking power law fall-offs, we can invert the above expression as an expansion in $1/r$ and demand that is only contain power law fall-offs, while comparing to \eqref{rtransformSch}. The result is 
\bea
f(v) = v + 4 m  \log v -\frac{\rho_{02} + 8m \log 2}{2}+\frac{\rho_{02}^2-2 \rho_{03}- 8 m \rho_{02} -64 m^2}{4 v} + \mathcal{O}\left(\frac{1}{v^2}\right)...\label{generalf}
\eea
We have set $\rho_{01}=2$. We can clearly expand to arbitrary high orders as required, but we are only retaining the non-trivial hair parameters ($\rho_{02}$ and $\rho_{03}$ in the notation here), on top of the mass.

Let us also note that \eqref{major} reduces to the Schmidt-Stewart form, if we simply declare that $f(v)=v$. The price we pay is that this leads to log fall-offs (including a piece that causes divergence in the charges) in the double null metric, if we compute $1/r$ from \eqref{major}. 
It should be clear from the form \eqref{generalf} that the key ingredient required to get rid of the log terms is the $ 4 m \log v$ on the RHS. The rest of the terms on the RHS simply change the power law coefficients, but do not change the fact that the series only contains power law fall-offs.

One simple choice for this coordinate transformation which is consistent with SDN gauge is to simply choose the $\rho_{ij}$ so that the higher order terms in $f(v)$ are all zero. This amounts to 
\bea
f(v)=v+4m \log v \label{simplest}
\eea
This only allows the Schwarzschild mass in the SDN form of metric. A slightly more general scenario that allows the two soft hair integration constants on top of the Schwarzschild mass is to truncate the series at $\mathcal{O}(1/v)$, and to choose the rest of the $\rho_{0i}$ to be zero (or choose them so that the higher order terms in $f(v)$ get fixed to zero\footnote{Since these higher terms correspond to trivial hair, both these choices correspond to the same physical solution.}.

To summarize: the expression \eqref{major} constitutes the general definition of a spherically symmetric vacuum solution of Einstein equations in the SDN gauge that we were after. Within the fall-offs considered in \cite{KP2}, the $f(v)$ contains up to three parameters which can be viewed as physical that can show up as non-trivial parameters in the covariant charge expressions, one combination of which is the usual Schwarzschild mass. The metric in terms of $f(v)$ can be written as 
\bea
ds^2&=&-2 \left(\frac{\partial r}{\partial v}\right) du\, d v + 2\, r^2\, \gamma_{z\zbr}\, dz\, d\zbr \\
&=& -\Big(1-\frac{2m}{r}\Big)\, \left(\frac{\partial f}{\partial v}\right) du\, d v + 2\, r^2\, \gamma_{z\zbr}\, dz\, d\zbr
\eea
It should of course be understood that $r=r(u,v)$ is an implicit function of $u$ and $v$ via the defining equation \eqref{major}. The minimal choice, which corresponds to ``pure'' Schwarzschild with no soft hair can be taken to be \eqref{simplest}. This yields
\bea
ds^2&=&-\Big(1-\frac{2m}{r}\Big)\left(1+\frac{4m}{v}\right) du\, d v + 2\, r^2\, \gamma_{z\zbr}\, dz\, d\zbr
\eea
This is an SDN form for conventional Schwarzschild (without any extra hair) that is asymptotically flat. The $r(u,v)$ here is the one defined via \eqref{major} with $f(v)$ from \eqref{simplest}. There exists an entirely analogous expression where $v \leftrightarrow -u$ that is valid around $\scrim$.

\section{Conclusions}

Let us make a few remarks about some of the (perhaps) surprising consequences of our results. 

\subsection{Time-Reversal, Time-Translation and Other Curiosities}

\begin{itemize}
\item We mentioned in Section \ref{SDNexp} that if we assume that the metric is only a function of $(v-u)$, then the only solution that has a power series expansion in $1/r \sim  1/(v-u)$ is the Minkowski metric. We saw a related fact, earlier: Schmidt-Stewart metric requires log fall-offs in $r$. This means that massive spacetimes need to be non-analytic at infinity in terms of the $r$ coordinate, a fact noticed in \cite{SS} as well as in many later papers. Note that if the metric can be written as a function of $(v-u)$, then it has a {\em manifest} CPT symmetry between the future and past which exchanges $v$ and $-u$ \cite{KP}.  Note that here we are talking about a CPT symmetry of the full $bulk$ metric. {\em Asymptotically} of course, CPT invariance is a natural demand between future and past SDN gauges for {\em any} solution \cite{KP} -- see analogous discussions in Bondi gauge in \cite{StromingerBMS}. 

\item One of the messages of this paper is that we can evade the above problem, if we allow power law fall-offs in $1/v$ at $\scrip$ (and $1/u$ at $\scrim$).  This allows us to have non-trivial spherically symmetric solutions, a  consequence of Section \ref{SDNexp}. Bulk CPT invariance is realized in these solutions through a {\em pair} of SDN charts (one each at $\scrip$ and $\scrim$). The two charts go to each other under $(v \leftrightarrow -u)$.  For Minkowski space, this is trivially true because both charts can be identified. This is less trivial for more general spherically symmetric solutions, but again quite automatic, via the $(u \leftrightarrow -v)$ replacement. Bulk CPT is therefore still a symmetry of the solutions we discussed in the previous sections -- it is simply that we need two charts to realize the solution\footnote{In other words, a massive spacetime needs a non-trivial representation of the $\IZ_2$ associated to CPT on charts.}. Note that a similar situation is true in Bondi gauge as well -- to see that the spacetime is time-reversal invariant ($u \leftrightarrow -v$), we need both the future and past Eddington-Finkelstein-Bondi charts.
as
\item Time reversal/CPT  implies that the spacetime is static, if the geometry has bulk time translation invariance (aka it is stationary). A natural way to implement time translations is to shift $v$ and $u$ by the {\em same} constant. This is an SDN supertranslation, see \cite{KP, KP1, KP2} and also Appendix \ref{asdiffs}. It can be seen from the Appendix that when one does this, the parameters $\mathscr{C}, \mathscr{D}^{(0)}$ and $\mathscr{D}^{(1)}$ in the general spherically symmetric solution in SDN gauge (Section \ref{SDNexp}) remain invariant, 
while further subleading metric parameters transform non-trivially. But since they are necessarily trivial hair parameters, changes in them are immaterial and we come to the conclusion that time translation invariance is realized in a pretty remarkable way in SDN gauge. Note that if the spherically symmetric metric was simply a function of $(v-u)$, time translation invariance and CPT invariance would have been manifest.

\item It is interesting to note that it is by doing an appropriate ``shift" in the definition of the diffeomorphisms (see \cite{KP1, KP2} and Appendix \ref{asdiffs}), that we are able to make the $subleading$ hypertranslation hair invariant under time translations as noted in the previous bullet point. This further emphasizes the observation in \cite{KP1, KP2} that the shifts basically correspond to separating out the independent diffeomorphisms. Doing the shifts was necessary to avoid mixing between the generators in the final algebra, and to put them into a ``diagonal" basis. There were some ambiguities in the choice of these shifts which were fixed in \cite{KP1, KP2} by considering the asymptotic Killing variations of the metric around the Riemann flat solution. The demand of time translation invariance in this paper leads us to a slightly different (and in hindsight, perhaps more natural) shift, which works with variations around Einstein solutions. Remarkably, the presentation of the algebra we found in \cite{KP2} remains unchanged despite the new form of the shift. This is gratifying, because the form of the algebra obtained in \cite{KP1, KP2} that extended the BMS algebra was quite a natural one.


\item The discussions of \cite{KP2} ensure that for the fall-offs considered there, leading hypertranslations are a global symmetry. The precise interpretation of {\em subleading} hypertranslations and whether they are a global symmetry or a trivial diffeomorphism, was not fully resolved in \cite{KP2}. The metric parameters affected by them showed up in the charges, but the diffeomorphisms themselves did not. One can check this for the charge expressions we have presented around Schwarzschild in Appendix \ref{charge} as well. These observations made their interpretation ambiguous because their status fell somewhere between global symmetries and trivial diffeomorphisms.


\item One reason to think that the subleading hypertranslation hair is trivial, is that the hair associated to them does not show up the in the vacuum Einstein constraints, see eqns (10) and (11) of \cite{KP}. Those equations were written down under the assumption of slightly more restrictive fall-offs than what we consider in \cite{KP2}, but this statement is true even with the more general constraints. This suggests that it may be reasonable to view the subleading hypertranslations as not part of radiative data. But at this stage, this is speculative\footnote{It is also worth investigating whether subleading hypertranslations can be made unambiguously physical, by considering fall-offs different from those in \cite{KP2}.}. A more clear understanding of the covariant surfaces charges and their interpretation is required before a reliable conclusion can be drawn about the status of subleading hypertranslations and their associated hair. We will not discuss this issue further here, and leave it as a future problem to unambiguously settle their status.

\item These caveats do not apply to (leading) hypertranslations. A set of fall-offs were identified in \cite{KP2} under which they are global symmetries. The observation of \cite{KP2} was that if one defines the vacuum state by Riemann flatness in SDN gauge, the solution can be more general that the conventional Minkowski form:
\beq 
ds^2=- du\, dv +2\, \Big(\frac{v-u}{2}\Big)^2\,  \gamma_{z\zbr}\, dz\, d\zbr 
\eeq
(Note that an identical philosophy leads one to supertranslation hair in Bondi gauge \cite{StromReview}.)  There exist \cite{KP2} fall-offs for which the charges are well-defined, and it is also easy to check that the Einstein constraints contain hypertranslation hair contributions. So if we want to claim that they are $not$ physical, we have to rule them out via more demands than those usually made in Birkhoff's theorem. As an aside, Riemann flatness in SDN gauge is a surprisingly rich idea, and we will have more to say about it elsewhere. We have considered the fall-offs of \cite{KP2} in this paper to illustrate our points, it will also be interesting to investigate whether there are other fall-offs that allow (other?) spherically symmetric soft hair.

\item Under a diffeomorphism, multiple metric functions are often affected. So what one calls hypertranslation hair is a bit of a convention. In this paper, we will often call $\mathscr{C}$ (or $C_{z\zbr}$ in the more general notation of \cite{KP2}) as the hypertranslation hair, but in some ways $\mathscr{C}+\mathscr{D}^{(1)}$ is also a natural choice. If one views $\mathscr{C}+\mathscr{D}^{(1)}$ and $\mathscr{C}-\mathscr{D}^{(1)}$ as the parameters of the metric, the former transforms under hypertranslations (see Appendix) but the latter do not, and has the interpretation as the SDN mass aspect. But $\mathscr{C}$ is the hypertranslation analogue of supertranslation shear in some ways, so we will retain this notation. 

\item If one declares that subleading hypertranslations are trivial, then the general spherically symmetric SDN metric we have identified in this paper has only two physical parameters ( $\mathscr{C}$ and $\mathscr{D}^{(1)}$), one combination of which is the mass. If instead they are non-trivial, we will have one more physical parameter, $\mathscr{D}^{(0)} $. In either case, the punchline remains that there exists fall-offs in SDN gauge which can evade Birkhoff's theorem. 

\item The above statement seems like a violation of Birkhoff's theorem, and in a sense it is. But this violation is only as dire as the fact that conventional soft hair modes in Bondi gauge violate the classic no-hair theorems of black holes. The underlying reason behind this is that soft modes are best thought of as a choice of diffeomorphism frame and how a black hole spacetime is ``attached'' to such a frame \cite{Flanagan}. Birkhoff's theorem is simply the oldest no-hair theorem, and our observation is just that in SDN gauge, we have some (previously unnoticed) choices in the spherically symmetric diffeomorphism frames.

\item We have worked with the vacuum Einstein equations in our papers so far, but in order to connect with memory effects and hard charges, it is useful to consider SDN gauge with a matter stress tensor. In particular, it may be interesting to understand the connection between hypertranslation hair and the wakes created by lightlike spherical shock waves \cite{Dray1, Dray2}. It is also conceivable that adding matter may lead to new consistency conditions (like energy positivity conditions) that may constrain or eliminate the {\em physical} mechanisms that can give rise to such memory effects. At the moment, these hair parameters simply label superselection sectors of vacuum gravity -- their physical origins and significance for gravitational waves (if any) remain to be understood. This will be discussed in \cite{future}, where the coupling to stress tensor will also be described.



\end{itemize}

\subsection{Summary}

Our goal in launching this paper was to obtain the Schwarzschild metric in the SDN gauge -- the usual double-null form of Schwarzschild due to Schmidt-Stewart \eqref{dnS} contains log fall-offs which lead to divergences in the charges. We have checked that the Schmidt-Stewart form of the metric contains a divergence in the charges at $\mathcal{O}(v \log v)$, due to the non-vanishing of the $\lambda_{01}$ term in the metric, see Appendix D eqn \eqref{loglambdaSch} for the notation $\lambda_{01}$. This is a divergence that is proportional to $\lambda_{01}$.

From the apparent consistency and generality of the power law fall-offs considered in \cite{KP, KP1}, it seemed inevitable that by restricting to the spherically symmetric case, one should be able to obtain Schwarzschild. But this leads only to an {\em asymptotic} definition. The asymptotic definition contained an infinite number of parameters on top of the Schwarzschild mass. Considering the fundamental significance of the Schwarzschild solution, one wishes for a definition that was non-perturbative in $\frac{1}{v}$ (or $\frac{1}{u}$ when thinking of $\scrim$). This turned out to be surprisingly tricky. In the Schmidt-Stewart case, the metric could be written as a function of $(u-v)$. One may be forgiven for assuming that the metric can be written as a function of $(u-v)$ even when the fall-offs are power law, because the action of asymptotic time translations is to shift $u$ and $v$ by equal amounts \cite{KP, KP1, KP2}, and conventional Birkhoff's theorem guarantees time translation invariance for the full metric. But we found that this was too naive an assumption. If we demand power law fall-offs in $(u-v)$ and spherical symmetry, Minkowski metric is the only allowed solution for Einstein. 

In this paper, we found that the existence of trivial asymptotic diffeomorphisms in SDN gauge \cite{KP, KP1, KP2} was crucial for resolving this puzzle. All of the integration constants, except the mass, hypertranslation hair and subleading hypertranslation hair turn out to be trivial. Hypertranslation hair is a soft mode associated to broken hypertranslation diffeomorphisms. The latter are global symmetries, whose covariant charges can be explicitly computed \cite{KP2}. It is known that soft hair can evade no hair theorems, the novelty in SDN gauge is simply that it allows spherically symmetric soft hair and therefore allows us to evade the oldest no-hair theorem, namely Birkhoff's theorem. 

The status of subleading hypertranslations is more ambiguous. The charge contains the corresponding hair parameters, but not the subleading hypertranslation transformations themselves. Genuine global symmetries contain both, while trivial diffeomorphisms contain neither. If one adopts the point of view that subleading hypertranslations are global symmetries, then these hair parameters should also be viewed as soft hair. But if instead they are trivial, then they can be ignored along with the rest of the (infinitely many) trivial hair parameters. A more refined study of covariant asymptotic charges is needed to fully settle this issue.

Thanks to these new features, bulk time translations are realized in an interesting way. The key point is that even though the form of the metric  does not remain invariant under a simultaneous shift of $u$ and $v$, it only changes up to {\em trivial} hair parameters. In other words the mass parameter, hypertranslation hair and subleading hypertranslation hair\footnote{The latter requires a ``shift" of the kind discussed in \cite{KP1, KP2} to unambiguously separate them from time translations. This raises the possibility that even in the more subleading (trivial) parameters, it may be possible to make the time translation invariance $manifest$, by doing shifts at all orders. We do not explore this possibility because this does not affect our conclusion regarding time translation invariance.} do remain invariant, even while the infinite number of trivial hair parameters transform. 


To conclude, Schwarzschild in the SDN gauge reveals many remarkable features which are inaccessible in other gauges. We identified a new soft spherically symmetric hair parameter, associated  to leading hypertranslations. This is a wrinkle in Birkhoff's theorem. The status of subleading hypertranslation hair is less clear, they may perhaps be trivial -- more work is required to resolve this issue. The point remains, however, that spherically symmetric asymptotic soft hair can be used to evade Birkhoff's theorem.


\section*{Acknowledgments}

We thank Glenn Barnich, Justin David and Karan Fernandes for discussions.  We also thank the audiences at various conferences and institutions (IISER Mohali, IIT Mandi, Arizona State University, University of Arizona and Caltech) where talks based on this material were presented during the last year -- for feedback, discussions and (numerous) questions.

\appendix

\section{Covariant Surface Charges Around Schwarzschild}\label{charge}

We will write down the charge expression around the general spherically symmetric solution here. Note that the background values are those of section \ref{SDNexp}, but the variations around it are arbitrary. The fall-offs we work with in this paper are those of \cite{KP2}, and the reader should consult that paper for notation. It can be checked that the general charge expression there reduces to the form below when the background is restricted to Schwarzschild. It can be seen that hypertranslations and their hair both show up in the charges, but subleading hypertranslations are not present (even though their hair is present). 
\beq\begin{aligned}
    \slashed{\delta}\mathcal{Q}_{\xi}[h;g] &=\frac{1}{16\pi G}\int d^2\Omega\, \Big[Y_A \Big(-\frac{1}{2}\, \mathcal{C}\, \delta\alpha^A_2-\frac{3}{8}\, \delta\alpha^A_3-\frac{1}{8}\, \partial_u\delta\alpha^A_4-\frac{1}{4}\, \mathcal{C}\, \partial_u\delta\alpha^A_3\Big)\\& +\psi\Big(-\frac{1}{4}\, \delta\lambda_2-\frac{1}{2}\, \gamma^{z\zbr}\, \delta\mathcal{D}_{z\zbr}+\frac{1}{4}\, D_A\delta\alpha^A_3+ \frac{1}{4}(\mathcal{C}+u) D_A\delta\alpha^A_2+\frac{1}{8}(\mathcal{C}-2u)\, \gamma^{z\zbr}\, \delta\mathcal{C}_{z\zbr}\Big) \\&+f \Big(-\frac{1}{2}D_A\delta\alpha^A_2-\frac{1}{2}\gamma^{z\zbr}\, \delta\mathcal{C}_{z\zbr}+\frac{1}{2}\gamma^{z\zbr}\, \partial_u\delta\mathcal{D}_{z\zbr}-\frac{1}{4}\partial_uD_A\delta\alpha^A_3\Big)-\frac{1}{2}\Delta_{\gamma}f\, D_A\delta\alpha^A_2\\&+X_A\Big(-\frac{1}{4}\delta\alpha^A_2-\frac{1}{8}\partial_u\delta\alpha^A_3-\frac{1}{4}D^AD_B\delta\alpha^B_2\Big)+\phi\Big(-\frac{1}{2}D_A\delta\alpha^A_2\Big)\Big]
\end{aligned}\eeq

\section{Asymptotic Diffeomorphisms and their Hair} \label{asdiffs}

General infinitesimal spherically symmetric diffeomorphisms can be captured by $u$ and $v$ dependent shifts in $u$ and $v$:
\bea
u \rightarrow u + \xi^u, \ \ v \rightarrow v+\xi^v
\eea
The demand that they satisfy the exact and asymptotic Killing vector conditions \cite{KP1, KP2} put constraints on their form. Here we discuss some features of these diffeomorphisms.

We write our Schwarzschild metric in SDN gauge as 
\beq ds^2 = -e^{\lambda(u,v)}\, du\, dv + 2\Big(\frac{v-u}{2}\Big)^2\, \Omega(u,v)\, dz\, d\zbr\eeq
where
\beq\begin{aligned}
\Omega(u,v) &= 1 + \frac{\mathscr{C}}{v} + \frac{\mathscr{D}^{(1)}\, u + \mathscr{D}^{(0)}}{v^2}+\mathcal{O}\big(1/v^3\big)\\
\lambda(u,v) &= \Big(\frac{1}{2}\big(\mathscr{C} - \mathscr{D}^{(1)}\big)\, u+\frac{1}{8}\big(\mathscr{C}^2-4\, \mathscr{D}^{(0)}\big)\Big)\frac{1}{v^2}+\mathcal{O}\big(1/v^3\big)
\end{aligned}\eeq
The most general spherically symmetric form for the AKVs $\xi^u$ and $\xi^v$ for our purposes is \cite{ KP2}:
\beq\begin{aligned}
    \xi^u &= T \\
    \xi^v &= T + \phi + \frac{1}{v}\Big[\tau + \frac{1}{2}\big(\mathscr{C} - \mathscr{D}^{(1)}\big)\, T\Big] + \mathcal{O}\big(1/v^2\big)    
\end{aligned}\eeq
Note that they are automatically $u$-independent up to this order. The further subleading diffs can be shown to be trivial. The above form arises from the restriction of the results in \cite{KP2} to spherical symmetry. A feature of the results in \cite{KP1, KP2} was that it was found convenient to introduce certain shifts in the definitions so that the different diffeomorphisms are separated (ie., do not mix with each other). This enabled the final AKV algebra to be ``diagonalized" into a nice form. The shifts are not completely unique and there is some ambiguity in their choices. In \cite{KP1, KP2} we found it convenient to do shifts that were inspired by the action of the AKVs on the Riemann flat (``vacuum") solution. For example, in eqn. (48) of \cite{KP2} all terms except the first two contributed to the shift -- compare it to eqn. (8) of the same paper. In principle the $T \lambda^2_1$ term could have also been included, but it was omitted simply because the $\lambda^2_1$ term is not Riemann flat. In the present paper, we have found that if one requires that our solution is time translation invariant (in particular, we want the subleading hypertranslation hair to not transform under time translations), it is natural to include the $T \lambda^2_1$ term in the shift. This corresponds to considering the AKV action on Einstein solutions and not just Riemann flat solutions. The term $\frac{1}{2}\big(\mathscr{C} - \mathscr{D}^{(1)}\big)$ that shifts the definition of the subleading hypertranslations above is precisely this $T \lambda^2_1$ term restricted to our present Schwarzschild solution. It should be emphasized that (remarkably) the algebra of the AKVs found in \cite{KP2} remains intact even with this modified shift. 

The $u$ and $v$ coordinates transform under theses diffs as
\beq\begin{aligned}
    u \rightarrow u' &= u + T \\
    v \rightarrow v' &= v + T + \phi + \frac{1}{v}\Big[\tau + \frac{1}{2}\big(\mathscr{C} - \mathscr{D}^{(1)}\big)\, T\Big] + \mathcal{O}\big(1/v^2\big)
\end{aligned}\eeq
One can check that this results in the following transformation laws for the integration constants:
\beq\begin{aligned}
    \mathscr{C} \rightarrow \mathscr{C}' &= \mathscr{C} +2\, \phi \\
    \mathscr{D}_{(1)} \rightarrow \mathscr{D}_{(1)}' &= \mathscr{D}_{(1)} +2\, \phi \\
    \mathscr{D}_{(0)} \rightarrow \mathscr{D}_{(0)}' &= \mathscr{D}_{(0)} + \mathscr{C}\, \phi + \phi^2 + 2\, \tau
\end{aligned}\eeq
These results can be obtained directly, or by restricting the results of \cite{KP2} to spherical symmetry. Note that to write down the complete transformation laws for the integration constants at subleading orders, we would need to include the subleading diffeomorphisms as well (sub-subleading hypertranslations, for example) in the expansions for $\xi^v$. One can also check that the coefficients in the $\lambda$ fall-offs are consistent with the above transformation laws. For example, in the case of $\lambda_2$, we have
\beq \lambda'_2(u) = \frac{1}{2}\big(\mathscr{C}' - \mathscr{D}_{(1)}'\big)\, u+\frac{1}{8}\big(\mathscr{C}'^2-4\, \mathscr{D}_{(0)}'\big) \eeq
The coefficient of the linear term becomes
\beq\begin{aligned}
    \frac{1}{2}\big(\mathscr{C}' - \mathscr{D}_{(1)}'\big) &= \frac{1}{2}\big[\big(\mathscr{C} + 2\, \phi\big) - \big(\mathscr{D}_{(1)} + 2\, \phi\big)\big] \\ &= \frac{1}{2}\big(\mathscr{C} - \mathscr{D}_{(1)}\big)
\end{aligned}\eeq
The coefficient of the constant term becomes 
\beq\begin{aligned}
    \frac{1}{8}\big(\mathscr{C}'^2-4\, \mathscr{D}_{(0)}'\big) &= \frac{1}{8}\Big[\big(\mathscr{C} + 2\phi\big)^2-4\, \Big(\mathscr{D}_{(0)} + \mathscr{C}\, \phi + \phi^2 + 2\, \tau \Big)\Big] \\
    &= \frac{1}{8}\big(\mathscr{C}^2 - 4\, \mathscr{D}_{(0)} - 8\, \tau\big)
\end{aligned}\eeq
It can be checked that these expressions exactly match the linear and constant terms respectively at $\mathcal{O}\big(1/v^2\big)$ in $e^{\lambda(u',v')}\, du'\, dv'$.\\
We can further understand the general transformations above by analyzing three special cases:
\begin{enumerate}
    \item Supertranslations only $(T = a, \phi = \tau = 0)$:\\
    In this case, the $u$ and $v$ coordinates transform as
    \beq\begin{aligned}
    u \rightarrow u' &= u + a \\
    v \rightarrow v' &= v + a + \frac{a}{2\, v}\big(\mathscr{C} - \mathscr{D}^{(1)}\big) + \mathcal{O}\big(1/v^2\big)
\end{aligned}\eeq
    and the integration constants transform as
    \beq\begin{aligned}
    \mathscr{C} \rightarrow \mathscr{C}' &= \mathscr{C}\\
    \mathscr{D}_{(1)} \rightarrow \mathscr{D}_{(1)}' &= \mathscr{D}_{(1)}\\
    \mathscr{D}_{(0)} \rightarrow \mathscr{D}_{(0)}' &= \mathscr{D}_{(0)}
    \end{aligned}\eeq
    Thus, we see that the integration constants are invariant under the action of constant supertranslations.
    \item Hypertranslations only $(\phi = a, T = \tau = 0)$:\\
    In this case, the $u$ and $v$ coordinates transform as
    \beq\begin{aligned}
    u \rightarrow u' &= u\\
    v \rightarrow v' &= v + a + \mathcal{O}\big(1/v^2\big)
    \end{aligned}\eeq
    and the integration constants transform as
    \beq\begin{aligned}
    \mathscr{C} \rightarrow \mathscr{C}' &= \mathscr{C} +2\, a \\
    \mathscr{D}_{(1)} \rightarrow \mathscr{D}_{(1)}' &= \mathscr{D}_{(1)} +2\, a \\
    \mathscr{D}_{(0)} \rightarrow \mathscr{D}_{(0)}' &= \mathscr{D}_{(0)} + \mathscr{C}\, a + a^2
    \end{aligned}\eeq
    \item Hypertranslations cancel supertranslations $\Big(T = a, \phi = -a, \tau = -\frac{1}{2}a\big(\mathscr{C} - \mathscr{D}^{(1)}\big)\Big)$:\\
    In this case, the $u$ and $v$ coordinates transform as
    \beq\begin{aligned}
    u \rightarrow u' &= u + a \\
    v \rightarrow v' &= v + \mathcal{O}\big(1/v^2\big)
    \end{aligned}\eeq
    and the integration constants transform as
    \beq\begin{aligned}
    \mathscr{C} \rightarrow \mathscr{C}' &= \mathscr{C} -2\, a \\
    \mathscr{D}_{(1)} \rightarrow \mathscr{D}_{(1)}' &= \mathscr{D}_{(1)} -2\, a \\
    \mathscr{D}_{(0)} \rightarrow \mathscr{D}_{(0)}' &= \mathscr{D}_{(0)} + \mathscr{D}_{(1)}\, a - 2\, \mathscr{C}\, a + a^2
\end{aligned}\eeq
\end{enumerate}

The ``supertranslations only" case, because we are working with the spherically symmetric situation and therefore dealing only with constant shifts, has the natural interpretation as a time translation. As promised earlier, we see that none of the integration constants transform under time translations, and the solution is therefore stationary. (Of course the further subleading trivial parameters may transform, but this does not affect our discussion.) 

\section{Einstein equations for Schwarzschild in SDN gauge} \label{SphericalEinstein}

The Einstein equations for the general spherically symmetric SDN metric 
\beq ds^2 = -e^{\lambda(u,v)}\, du\, dv + 2\Big(\frac{v-u}{2}\Big)^2\, \Omega(u,v)\, dz\, d\zbr\eeq
yield the following set of second-order PDEs:\\

\noindent
Setting the $uu-$component of the Einstein tensor to vanish gives
\beq \partial_u^2\Omega-\frac{1}{2}\frac{(\partial_u\Omega)^2}{\Omega} - (\partial_u\Omega)(\partial_u\lambda) - \frac{2}{v-u}(\partial_u\Omega) + \frac{2}{v-u}\Omega\, (\partial_u\lambda) = 0 \eeq
Setting the $vv-$component of the Einstein tensor to vanish gives
\beq \partial_v^2\Omega-\frac{1}{2}\frac{(\partial_v\Omega)^2}{\Omega} - (\partial_v\Omega)(\partial_v\lambda) + \frac{2}{v-u}(\partial_v\Omega) - \frac{2}{v-u}\Omega\, (\partial_v\lambda) = 0 \eeq
Setting the $uv-$component of the Einstein tensor to vanish gives
\beq\label{uv} \partial_u\partial_v\Omega + \frac{2}{v-u}(\partial_u\Omega - \partial_v\Omega) + \frac{2}{(v-u)^2}(e^{\lambda} - \Omega) = 0 \eeq
Setting the $z\zbr-$component of the Einstein tensor to vanish gives
\beq\label{zzbr} \partial_u\partial_v\lambda + \frac{\partial_u\partial_v\Omega}{\Omega} + \frac{1}{v-u}\frac{(\partial_u\Omega - \partial_v\Omega)}{\Omega} - \frac{1}{2} \frac{(\partial_u\Omega)(\partial_v\Omega)}{\Omega^2} = 0 \eeq
Substituting \eqref{uv} in \eqref{zzbr} gives
\beq \partial_u\partial_v\lambda - \frac{1}{2}\frac{(\partial_u\Omega)(\partial_v\Omega)}{\Omega^2} - \frac{1}{v-u}\frac{(\partial_u\Omega - \partial_v\Omega)}{\Omega} - \frac{2}{(v-u)^2}\Big(\frac{e^{\lambda}}{\Omega}-1\Big) = 0 \eeq

As mentioned in the main text, we have investigated the possibility that the metric is only a function of $(v-u)$ in the SDN form. This amounts to
\beq\begin{aligned}
    \Omega(u, v) &= \Omega\Big(\frac{v-u}{2}\Big) \equiv \Omega(r) \\ 
    \lambda(u, v) &= \lambda\Big(\frac{v-u}{2}\Big) \equiv \lambda(r)
\end{aligned}\eeq
where we have defined $r = \frac{v-u}{2}$. This gives us the following results:
\beq 
    \partial_u\Omega = -\frac{1}{2}\partial_r\Omega \qquad
    \partial_v\Omega = \frac{1}{2}\partial_r\Omega \qquad
    \partial_u\partial_v\Omega = -\frac{1}{4}\partial_r^2\Omega \qquad
    \partial_u^2\Omega = \frac{1}{4}\partial_r^2\Omega \qquad
    \partial_v^2\Omega = \frac{1}{4}\partial_r^2\Omega
\eeq
along with similar results for $\lambda$. Thus the above set of PDEs can be rewritten as the following set of ODEs:
\beq\begin{aligned}
    & \Omega'' - \frac{\Omega'^2}{\Omega} - \Omega'\, \lambda' + \frac{2}{r}\, \Omega' - \frac{2\, \Omega}{r}\, \lambda' = 0 \\
    & \Omega'' + \frac{4}{r}\, \Omega' - \frac{2}{r^2}\, (e^{\lambda} - \Omega) = 0 \\
    & \lambda'' - \frac{1}{2}\, \Big(\frac{\Omega'}{\Omega}\Big)^2 - \frac{4}{r}\, \Big(\frac{\Omega'}{\Omega}\Big) + \frac{2}{r^2}\, \Big(\frac{e^{\lambda}}{\Omega}-1\Big) = 0
\end{aligned}\eeq
where the primes denote derivatives with respect to $r$. Let us emphasize that here one should think of $r$ as simply a proxy for the variable $\frac{v-u}{2}$, and we are working still in the double null coordinate system.  
By expanding $\Omega(r)$ and $\lambda(r)$ in powers of $1/r$, one can show that $\lambda = 0$, $\Omega = 1$ is the only possible solution. Thus by demanding that $\lambda(u, v)$ and $\Omega(u, v)$ are only functions of $(v-u)$, and demanding power law fall offs in $1/r$, we are lead to Minkowski space as the unique solution. Note that this is consistent with our earlier observation that the Schmidt-Stewart metric for Schwarzschild, which is a double null coordinate system and is only a function of $(v-u)$, contains log fall-offs.

\section{Schwarzschild in SDN Gauge: With Log Fall-Offs}\label{logs}

In this and the next Appendices, we will determine the general spherically symmetric SDN gauge solution asymptotically, by assuming that the fall-offs contain power laws in $1/v$ and also $\log v$. The analogous problem where only power laws were considered, was discussed in Section \ref{SDNexp}.  This discussion will enable us to connect with the discussion of the Schmidt-Stewart metric, which contains log terms. 

We begin as usual with the spherically symmetric double null gauge metric written as follows
\beq ds^2=-e^{\lambda(u,v)}\, du\, dv +2\, \Big(\frac{v-u}{2}\Big)^2\, \Omega(u,v)\, \gamma_{z\zbr}\, dz\, d\zbr \eeq
while considering the asymptotic fall-off conditions for the gauge functions $\lambda(u,v)$ and $\Omega_{z\zbr}(u,v)$ written using the expansion parameter $v$ as

\beq\label{loglambdaSch}\begin{aligned}&\lambda(u,v) = \bigg(\frac{\lambda_{01}(u)}{v}+\frac{\lambda_{02}(u)}{v^2}+\frac{\lambda_{03}(u)}{v^3}+O\big(v^{-4}\big)\bigg)\\&+(\ln{v})\bigg(\frac{\lambda_{12}(u)}{v^2}+\frac{\lambda_{13}(u)}{v^3}+O\big(v^{-4}\big)\bigg)+(\ln{v})^2\bigg(\frac{\lambda_{23}(u)}{v^3}+O\big(v^{-4}\big)\bigg)+O\big[(\ln{v})^3\big]\end{aligned}\eeq

\beq\label{logomegaSch}\begin{aligned}
&\Omega_{z\zbr}(u,v)= \bigg(1+\frac{\mathcal{C}^{(0)}(u)}{v}+\frac{\mathcal{D}^{(0)}(u)}{v^2}+\frac{\mathcal{E}^{(0)}(u)}{v^3}+O\big(v^{-4}\big)\bigg)\\& +(\ln{v})\bigg(\frac{\mathcal{C}^{(1)}(u)}{v}+\frac{\mathcal{D}^{(1)}(u)}{v^2}+\frac{\mathcal{E}^{(1)}(u)}{v^3}+O\big(v^{-4}\big)\bigg)+(\ln{v})^2\bigg(\frac{\mathcal{D}^{(2)}(u)}{v^2}+\frac{\mathcal{E}^{(2)}(u)}{v^3}+O\big(v^{-4}\big)\bigg)\\&+(\ln{v})^3\bigg(\frac{\mathcal{E}^{(3)}(u)}{v^3}+O\big(v^{-4}\big)\bigg)+O\big[(\ln{v})^4\big]\end{aligned}\eeq

Next, we would like to demand that the spherical symmetric double null gauge metric along with the fall-offs specified in \eqref{loglambdaSch}-\eqref{logomegaSch} satisfies the Einstein's equations in vacuum.
\begin{itemize}
    \item Constraints on $\Omega$ fall-offs in \eqref{logomegaSch}:
    \begin{subequations}
        \begin{align}
        \partial_u\mathcal{C}^{(1)} &= 0 \implies \mathcal{C}^{(1)} = \text{constant} = \mathscr{C}^{(1)}\\
        \partial_u\mathcal{C}^{(0)} &= 0 \implies \mathcal{C}^{(0)} = \text{constant} = \mathscr{C}^{(0)}\\
        \partial_u\mathcal{D}^{(2)} &= 0 \implies \mathcal{D}^{(2)} = \text{constant} = \mathscr{D}^{(2)}, \quad \mathscr{D}^{(2)} = \frac{1}{4} \big(\mathscr{C}^{(1)}\big)^2\\
        \partial_u\mathcal{D}^{(1)} &= \mathscr{C}^{(1)} \implies \mathcal{D}^{(1)} = \mathscr{C}^{(1)}\, u+ \mathscr{D}^{(1)}\\
        \partial_u^2\mathcal{D}^{(0)} &= 0 \implies \mathcal{D}^{(0)} = \mathscr{D}^{(0a)}\, u + \mathscr{D}^{(0b)}\\
        \partial_u\mathcal{E}^{(3)} &= 0 \implies \mathcal{E}^{(3)} = \text{constant} = \mathscr{E}^{(3)}, \quad \mathscr{E}^{(3)}=0\\
        \partial_u\mathcal{E}^{(2)} &= \frac{1}{2}\big(\mathscr{C}^{(1)}\big)^2 \implies \mathcal{E}^{(2)} = \frac{1}{2}\big(\mathscr{C}^{(1)}\big)^2\, u+ \mathscr{E}^{(2)}\\
        \partial_u\mathcal{E}^{(1)} &= 2\mathscr{C}^{(1)}\, u+\frac{1}{2}\Big(2\mathscr{D}^{(1)}+\mathscr{C}^{(1)}\mathscr{C}^{(0)}\Big) \implies \mathcal{E}^{(1)} = \mathscr{C}^{(1)}\, u^2+\frac{u}{2}\Big(2\mathscr{D}^{(1)}+\mathscr{C}^{(1)}\mathscr{C}^{(0)}\Big)+ \mathscr{E}^{(1)}\\
        \begin{split}
        \partial_u\mathcal{E}^{(0)} &= \big(3\, \mathscr{D}^{(0a)}-\mathscr{C}^{(1)}\big)\, u+\Big(\mathscr{D}^{(0b)}+\frac{1}{4}\, \big(\mathscr{C}^{(1)}\big)^2\Big) \\& \implies \mathcal{E}^{(0)} = \frac{1}{2}\, \big(3\, \mathscr{D}^{(0a)}-\mathscr{C}^{(1)}\big)\, u^2+ \Big(\mathscr{D}^{(0b)}+\frac{1}{4}\, \big(\mathscr{C}^{(1)}\big)^2\Big)\, u+\mathscr{E}^{(0)}
        \end{split}
        \end{align}
    \end{subequations}    
    \item Constraints on $\lambda$ fall-offs in \eqref{loglambdaSch}:
    \begin{subequations}
        \begin{align}
        \lambda_{01} &= \frac{1}{2}\mathscr{C}^{(1)}\\
        \lambda_{12} &= \frac{1}{4}\big(\mathscr{C}^{(0)}\mathscr{C}^{(1)}-2\mathscr{D}^{(1)}\big)\\
        \lambda_{02} &= \frac{1}{2}\, \big(\mathscr{C}^{(0)}-\mathscr{D}^{(0a)}\big)\, u + \frac{1}{8}\, \Big(\big(\mathscr{C}^{(0)}\big)^2-4\, \mathscr{D}^{(0b)}-2\mathscr{C}^{(0)}\mathscr{C}^{(1)}-\big(\mathscr{C}^{(1)}\big)^2+4\mathscr{D}^{(1)}\Big)\\
        \lambda_{23} &= -\frac{1}{4}\, \mathscr{C}^{(1)}\, \big(\mathscr{C}^{(0)}\mathscr{C}^{(1)}-2\mathscr{D}^{(1)}\big)-\mathscr{E}^{(2)}\\
        \begin{split}
            \lambda_{13} &= \frac{1}{2}\big(\mathscr{D}^{(0a)}\mathscr{C}^{(1)}-\mathscr{C}^{(0)}\mathscr{C}^{(1)}\big)u+\frac{1}{8}\Big(-3\big(\mathscr{C}^{(0)}\big)^2\mathscr{C}^{(1)}+4\mathscr{D}^{(0b)}\mathscr{C}^{(1)}+\mathscr{C}^{(0)}\big(\mathscr{C}^{(1)}\big)^2\\&+4\mathscr{C}^{(0)}\mathscr{D}^{(1)}-2\mathscr{C}^{(1)}\mathscr{D}^{(1)}-8\, \mathscr{E}^{(1)}+8\, \mathscr{E}^{(2)}\Big)
        \end{split}\\
        \begin{split}
            \lambda_{03} &= \frac{1}{2}\, \big(\mathscr{C}^{(0)}-\mathscr{D}^{(0a)}\big)-\frac{1}{2}\, \mathscr{C}^{(0)}\, \big(\mathscr{C}^{(0)}-\mathscr{D}^{(0a)}\big)+\frac{1}{24}\Big(-3\big(\mathscr{C}^{(0)}\big)^3+12\, \mathscr{C}^{(0)}\, \mathscr{D}^{(0b)}\\&-24\, \mathscr{E}^{(0)}+3\, \big(\mathscr{C}^{(0)}\big)^2 \mathscr{C}^{(1)}+3\, \mathscr{C}^{(0)}\big(\mathscr{C}^{(1)}\big)^2+\big(\mathscr{C}^{(1)}\big)^3-6\, \mathscr{C}^{(0)}\mathscr{D}^{(1)}-6\, \mathscr{C}^{(1)}\mathscr{D}^{(1)}+12\, \mathscr{E}^{(1)}\Big)
        \end{split}
        \end{align}
    \end{subequations}
\end{itemize}

This solution provides the extension of the discussion in Section \ref{SDNexp} that contains log fall-offs. Our interest in this solution has to do with the fact that the Schmidt-Stewart metric has log terms in it. We will clarify the connection in the next Appendix.

\section{Schwarzschild in SDN Gauge with Log Fall-Offs: Via Bondi} \label{logdiffs}

In this Appendix, we will show how the integration constants we found in the Einstein solutions are related to the parameters that fix the coordinate transformation \eqref{rtransformSch} from Schwarzschild.  Our strategy for carrying out the comparison between the two sets of results above will be to compare specific metric components, order by order, after doing the coordinate transformation. The reader should consult Section \ref{viaBondi} to see the definition of the $\rho_{ij}$ parameters, which are constants. They fix the coordinate transformation \eqref{rtransformSch} completely.

\begin{itemize}
    \item Constraints from comparing $g^{uv}$:
    \begin{subequations}
    \begin{align}
        \rho_{01} &= 2\\
        \lambda_{01} &= -\frac{1}{2}\rho_{12}\\
        \rho_{23} &= \frac{1}{2}\big(\mathscr{C}^{(1)}\big)^2\\
        \lambda_{12} &= \frac{1}{2}\big(\rho_{12}^2+\rho_{13}-2\rho_{23}-\rho_{02}\, \rho_{12}\big)\\
        \lambda_{02} &= -4\, M\, u+\frac{1}{8}\big(4\, \rho_{03}-2(\rho_{02})^2+4\, \rho_{02}\, \rho_{12}-(\rho_{12})^2-4\, \rho_{13}\big)\\
        \rho_{34} &= -\frac{1}{4}\big(\mathscr{C}^{(1)}\big)^3\\
        \lambda_{23} &= \frac{3}{4}\rho_{02}(\rho_{12})^2-\frac{5}{8}(\rho_{12})^3-\rho_{12}\, \rho_{13}-\rho_{02}\, \rho_{23}+2\, \rho_{12}\, \rho_{23}+\rho_{24}-\frac{3}{2}\rho_{34}\\
        \begin{split}
            \lambda_{13} &= \frac{3}{4}\rho_{12}(\rho_{02})^2+\rho_{14}-\rho_{02}\big(\rho_{12}^2+\rho_{13}-\rho_{23}\big)-\rho_{24}+\frac{1}{4}\rho_{12}\big((\rho_{12})^2-4\rho_{03}\\&+5\rho_{13}-2(\rho_{23}+8\, M\, u)\big)
        \end{split}
    \end{align}
    \end{subequations}
    \item Constraints from comparing $g^{z\zbr}$:
    \begin{subequations}
    \begin{align}
        \mathscr{C}^{(1)} &= -\rho_{12}\\
        \mathscr{C}^{(0)} &= -\rho_{02}\\
        \mathscr{D}^{(2)} &= \frac{3}{4}(\rho_{12})^2-\rho_{23}\\
        \mathscr{D}^{(1)} &= \frac{3}{2}\rho_{02}\rho_{12}-\rho_{13}\\
        \begin{split}
        \mathcal{D} &= \big(8\, m-\rho_{02}\big)\, u+\Big(\frac{3}{4}\, (\rho_{02})^2-\rho_{03}\Big)\\&
        \implies \mathscr{D}^{(0a)} =8\, m-\rho_{02}\\&
        \implies \mathscr{D}^{(0b)} =\frac{3}{4}\, (\rho_{02})^2-\rho_{03}
        \end{split}\\
        \mathscr{E}^{(3)} &= \frac{1}{2}\big(-(\rho_{12})^3+3\rho_{12}\rho_{23}-2\rho_{34}\big)\\
        \mathscr{E}^{(2)} &= \frac{1}{2}\Big(3\rho_{12}\rho_{13}+3\rho_{02}\big(\rho_{23}-(\rho_{12})^2\big)-2\rho_{24}\Big)\\
        \mathscr{E}^{(1)} &= \frac{1}{2}\big(3\rho_{03}\rho_{12}-3(\rho_{02})^2\rho_{12}+3\rho_{02}\rho_{13}-2\rho_{14}\big)
    \end{align}
    \end{subequations}    
\end{itemize}

The key observation here is that turning on the logarithmic term $r_{12}$ in the coordinate transformation (or equivalently $\rho_{12}$) turns on the $\lambda_{01}$ term in the metric. The presence of this term leads to a divergence in the charges and this is one way to see that the Schmidt-Stewart metric is not asymptotically flat.

\end{document}